\documentclass{article}
\usepackage{arxiv}
\usepackage{cite}
\usepackage{amsmath,amssymb,amsfonts}
\usepackage{algorithmic}
\usepackage{graphicx}
\usepackage{textcomp}
\usepackage{xcolor}
\usepackage{booktabs}
\usepackage{bm}
\usepackage{verbatim}
\usepackage{url}

\def\BibTeX{{\rm B\kern-.05em{\sc i\kern-.025em b}\kern-.08em
    T\kern-.1667em\lower.7ex\hbox{E}\kern-.125emX}}
\renewcommand{\shorttitle}{Towards Hierarchical Cyber Defense with Large Language Models}
\date{}
\begin{document}

\title{Towards Hierarchical Cyber Defense with Large Language Models: From Planning to Execution}

\author{Harshith Doppalapudi \\
\textit{Indiana University} \\
Bloomington, IN, USA \\
hdoppala@iu.edu
\And
Nathaniel D. Bastian \\
\textit{Johns Hopkins University} \\
Baltimore, MD, USA \\
ndbastian@jhu.edu
\And
Ankit Shah \footnote{}\\
\textit{Indiana University} \\
Bloomington, IN, USA \\
ankit@iu.edu
}

\maketitle
{\renewcommand{\thefootnote}{}\footnotetext{Send correspondence to ankit@iu.edu \\ \copyright~2026 IEEE. Personal use of this material is permitted. Permission from IEEE must be obtained for all other uses, in any current or future media, including reprinting/republishing this material for advertising or promotional purposes, creating new collective works, for resale or redistribution to servers or lists, or reuse of any copyrighted component of this work in other works.}}

\begin{abstract}
An autonomous cyber defender trained with reinforcement learning (RL) is typically tied to the network on which it was trained, limiting its ability to generalize as network scale changes. Hierarchical RL reduces decision complexity by separating strategic targeting from tactical execution, but it does not eliminate this retraining dependence. We investigate whether frozen, zero-shot large language models (LLMs) can provide retraining-free control in hierarchical cyber defense and how performance changes as LLM control is extended from planning to execution. We formulate a controller-agnostic planner-executor hierarchy in which the planner selects a subnet to defend over a fixed horizon and the executor selects defensive actions within that subnet. Using the high-fidelity Cyberwheel environment, with its built-in automated red team agent mapped to the MITRE ATT\&CK framework, we compare RL+RL, LLM+RL, and LLM+LLM configurations using six models ranging from 3B to 70B parameters, including two cybersecurity-specialized models, across small, medium, and large networks. Replacing only the planner with an LLM yields limited gains as network size increases. In contrast, extending LLM control to execution produces notable improvements for sufficiently capable models. For instance, a frozen general purpose 70B model holds successful lateral movement to approximately 1\% of steps and attacker impact near zero across all three network scales using the same model weights, while the RL baseline is retrained for each scale. %Our results show that frozen LLMs can enable retraining-free, scale-generalizing hierarchical cyber defense, while also indicating that strong tactical execution is critical to realizing the benefits of LLM-based control.
Our results show that \textcolor{black}{sufficiently capable frozen LLMs can maintain strong defensive performance across the evaluated network scales without task-specific retraining}, while also indicating that strong tactical execution is important to realizing the benefits of LLM-based control.
\end{abstract}


\section{Introduction}
Autonomous cyber defense requires agents to make sequential decisions as an attacker moves through a network toward critical assets. As enterprise networks grow, this problem becomes increasingly difficult as the defender must determine both \emph{where} to intervene and \emph{which} countermeasure to execute, while operating from incomplete and noisy observations. Reinforcement learning (RL) provides a natural framework for learning such policies, but learned defenders are typically coupled to the network configurations on which they are trained. Changes in network size or topology can therefore require costly retraining, limiting their ability to transfer across operational environments.

Hierarchical reinforcement learning (HRL) addresses part of this challenge by decomposing defense into decisions at different levels of abstraction. A high level policy can select the region of the network requiring attention, while a lower-level policy determines the defensive action within that region~\cite{singh2025hier}. This decomposition reduces the complexity of the joint decision space and has shown promise for scalable autonomous cyber defense. However, hierarchy alone does not eliminate the dependence of learned policies on their training environments. A defender that must be retrained whenever the underlying network changes remains difficult to deploy across heterogeneous or evolving networks.

Pretrained large language models (LLMs) offer a fundamentally different form of generalization. A frozen LLM can exploit knowledge acquired during pretraining without task-specific parameter updates and can reason over structured representations of previously unseen situations. Recent studies have explored LLMs as autonomous cyber defenders ~\cite{llmdefenders} and as mechanisms for designing rewards for RL-based defenders~\cite{llmreward}. \textcolor{black}{However, prior work has not examined frozen LLM control within a two-level planner-executor hierarchy across vastly different network scales, nor how performance changes as LLM control is extended from strategic planning to tactical execution.}
%Recent studies have explored LLMs as autonomous cyber defenders~\cite{llmdefenders} and as mechanisms for designing rewards for RL-based defenders~\cite{llmreward}. These studies, however, largely consider flat controllers in small, fixed environments. It remains unclear whether pretrained LLMs can provide retraining-free control across significantly different network scales and, critically, where such capability should be introduced within a hierarchical defender.

The planner appears to be a natural fit. In a hierarchical network defender, the planner selects among a relatively small set of network regions, whereas the executor must translate that strategic choice into host-level defensive actions. Because the planner's decision space changes little as the underlying network grows, one might expect that replacing a learned planner with a frozen LLM would capture much of the generalization benefit while retaining an efficient RL executor. Yet this assumption has not been systematically tested. Also, it is unknown whether the principal limitation of a hierarchical defender lies in strategic targeting, i.e., deciding where to act, or in tactical execution, i.e., deciding which action to take once the target region has been selected.

We investigate these questions through a controller-agnostic planner-executor hierarchy in which either level can be instantiated by a frozen, zero-shot LLM. This design enables a controlled comparison of three architectures under identical observations, actions, rewards, attacker behavior, and hierarchical interfaces. These architectures include a fully trained RL hierarchy (RL+RL), an LLM planner coupled with a trained RL executor (LLM+RL), and an all-LLM hierarchy (LLM+LLM). We evaluate six general purpose and cybersecurity-specialized models ranging from 3B to 70B parameters in Cyberwheel~\cite{cyberwheel} across three network sizes (small, medium, and large).

Our experiments suggest a result that is more nuanced than simply replacing RL with an LLM. Introducing an LLM only at the planning level provides limited improvement as network size grows, and increasingly capable planners produce relatively small gains when paired with the same trained executor. In contrast, when a sufficiently capable frozen LLM controls both planning and execution, performance changes dramatically. For instance, we found that Llama-3.3-70B reduces and maintains compromise rate to approximately 1\% across all three network sizes and drives attacker impact to nearly zero, using the same model weights without retraining. This effect is not universal as smaller models and some cybersecurity-specialized models fail to maintain this robustness, indicating that pretrained control generalizes only when sufficient capability extends to the tactical layer. These patterns are directly reflected in our cross-scale RL+RL, LLM+RL, and LLM+LLM experimental results.

We therefore ask the following research questions (RQs):
\begin{itemize}
    \item \textbf{RQ1:} Can a frozen, zero-shot LLM-based hierarchical defender maintain effective cyber defense across network scales without task-specific retraining?

\item \textbf{RQ2:}
\textcolor{black}{How does extending LLM control from strategic planning to tactical execution affect hierarchical cyber defense across network scales?}
%    \item \textbf{RQ2:} Where in the hierarchy does LLM capability matter most for cross-scale defense: strategic planning or tactical execution?

    \item \textbf{RQ3:} How do model scale and cybersecurity specialization affect the robustness of frozen LLM controllers as network size increases?
\end{itemize}

Our study makes three contributions:

\begin{itemize}
    \item \textbf{Controller-agnostic hierarchical framework:} We develop a planner-executor architecture that permits frozen LLMs to be interchanged at the planner and executor levels while preserving the same environment, reward, observation, and action interfaces.

    \item \textbf{Controlled cross-scale evaluation:} We compare RL+RL, LLM+RL, and LLM+LLM defenders using six models from 3B to 70B parameters across a 67-fold increase in network size, explicitly testing whether pretrained controllers can avoid environment-specific retraining.

\item \textbf{\textcolor{black}{Characterization of LLM control across the hierarchy:}} We show that stronger planning alone provides limited benefit when tactical execution remains fixed, whereas extending sufficiently capable LLM control to execution yields substantially stronger cross-scale performance. We further find that cybersecurity specialization can improve particular models but does not, by itself, guarantee robustness as the network scales.
    %\item \textbf{Characterization of the hierarchical capability bottleneck:} We show that stronger planning alone provides limited benefit when tactical execution remains fixed, whereas sufficiently capable LLM control at both levels enables strong cross-scale generalization. We further find that cybersecurity specialization can improve particular models but does not, by itself, guarantee robustness as the network scales.
\end{itemize}

The rest of the paper is organized as follows. We first review the related work, then describe the hierarchical control architecture and experimental setup, and finally present the results, conclusions, and directions for future research.

\section{Related Work}
\label{sec:related}
\textbf{Autonomous cyber defense environments.}
Cyber defense gyms formulate blue-versus-red interaction as a sequential decision problem. CybORG and the associated CAGE challenges support increasingly complex autonomous defense scenarios, from earlier single agent settings ~\cite{autonomousagents} to CAGE Challenge~4, a scalable multi-agent environment in which defenders protect different network subnets~\cite{cage4}. Cyberwheel, used in this study, is a configurable simulation and emulation platform developed by Oak Ridge National Laboratory for evaluating autonomous cyber defense under controllable network, attacker, observation, action, and reward settings~\cite{cyberwheel}. Its configurability enables our controlled evaluation across network scales.

\textbf{Hierarchical reinforcement learning for cyber defense.}
Hierarchical RL reduces control complexity by decomposing high-level coordination from lower-level action selection. A hierarchical proximal policy optimization (PPO) architecture was recently proposed for cyber network defense in which a master policy coordinates PPO-trained sub-policies, addressing large policy spaces and improving performance and convergence across multiple adversaries~\cite{singh2025hier}. Their results also demonstrate that learned sub-policies can be transferred and fine-tuned when network or adversary conditions change. Building on this hierarchical control paradigm, we use PPO for the learned controllers and investigate whether frozen LLMs can provide effective control across network scales without task-specific retraining.

\textbf{Large language models for cyber defense.}
On the defensive side, recent work has evaluated LLM-based defenders in the multi-agent CybORG CAGE~4 environment, including mixed teams of LLM- and RL-based agents~\cite{llmdefenders}, while other studies have used LLM-generated reward functions to train RL-based defenders in Cyberwheel~\cite{llmreward}. These studies establish complementary roles for LLMs in autonomous cyber defense, but do not study frozen LLM control within a two-level planner-executor hierarchy across network scales. To our knowledge, prior work has not systematically compared planner-only LLM integration with extending LLM control to tactical execution under common hierarchical interfaces, nor evaluated the same frozen controllers across substantially different network sizes. We address these gaps through a controlled comparison of RL+RL, LLM+RL, and LLM+LLM while varying model scale and cybersecurity specialization.

\section{A Controller-Agnostic Hierarchical Control Architecture}
\label{sec:framework}
\begin{figure}[t]
\centering
\includegraphics[width=0.36\linewidth]{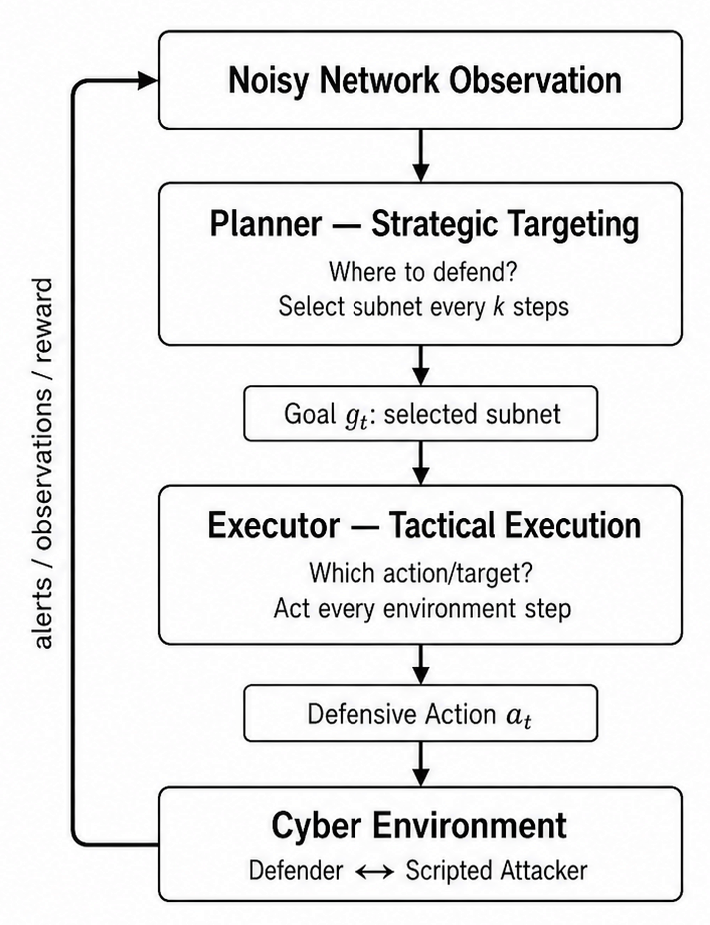}
\caption{Controller-agnostic planner-executor architecture. LLM+RL introduces frozen LLM control at strategic planning, while LLM+LLM extends it to tactical execution; RL+RL serves as the trained hierarchical baseline.}
\label{fig:framework}
\end{figure}

To separate the effect of controller type from its role in the hierarchy, we formulate autonomous cyber defense as a controller-agnostic planner-executor architecture. The planner determines \emph{where} defensive effort should be focused, while the executor determines \emph{which} action should be taken within the selected region. As illustrated in Figure~\ref{fig:framework}, this formulation supports a controlled progression from a fully learned hierarchy (RL+RL) to LLM-based planning (LLM+RL) and finally, to LLM control at both planning and execution levels (LLM+LLM). This design allows us to examine whether LLM-based planning alone is sufficient and what changes when LLM control is extended to tactical execution. The details about the environment, attacker, controller settings, and hyperparameters are described in Section~\ref{sec:setup}.

\subsection{Defense as a Partially Observable Decision Process}
We model the defender as a partially observable Markov decision process (POMDP)
\begin{equation}
\mathcal{M} =
\left(\mathcal{S},\mathcal{A},P,R,\Omega,O,\gamma\right),
\end{equation}
where $\mathcal{S}$ is the hidden environment state, $\mathcal{A}$ is the defender action set, $P(s'\mid s,a)$ is the transition kernel induced jointly by the defender and the scripted attacker, $R:\mathcal{S}\times\mathcal{A}\times\mathcal{S}\rightarrow\mathbb{R}$ is the reward, $\Omega$ is the observation space, $O(o\mid s)$ is the stochastic observation kernel, and $\gamma\in(0,1)$ is the discount factor. The hidden state captures network topology, per-host compromise, attacker kill-chain progress, and defender modifications.

The defender observes only noisy detector-derived evidence, $o_t\sim O(\cdot\mid s_t)$, rather than the true state. Because the observation includes decayed alert-history features (Section~\ref{sec:setup-env}), we use reactive policies conditioned on the current observation as a practical approximation to full history- or belief-state policies.

\subsection{Planner-Executor Decomposition}
Let the network be partitioned into $B$ subnets, defining the goal set
$\mathcal{G}=\{1,\ldots,B\}$, where each goal identifies the subnet on which the
defender should focus. A primitive defender action is
\begin{equation}
a=(\tau,u), \qquad u\in\mathcal{U},
\end{equation}
where $u$ is a defensive primitive and $\tau$ is its target when required. The target granularity depends on the primitive, which includes host-level for isolation, subnet-level for decoy placement, and none for a no-op.

The hierarchy separates two qualitatively different decisions. The \emph{planner} $\pi_{\mathrm{hi}}$ performs strategic targeting by selecting a subnet, whereas the \emph{executor} $\pi_{\mathrm{lo}}$ performs tactical execution by selecting the defensive action within that subnet. This distinction is central to our experimental design as the planner operates over the comparatively compact goal set $\mathcal{G}$, while the executor translates that goal into local host- and action-level decisions.

The planner acts every $k$ environment steps,
\begin{equation}
g_j \sim
\pi_{\mathrm{hi}}\!\left(\cdot\mid o^{\mathrm{hi}}_{jk}\right),
\qquad
g_t=g_{\lfloor t/k\rfloor},
\label{eq:planner}
\end{equation}
where $o^{\mathrm{hi}}$ is a subnet-level representation of the observation. The selected goal remains fixed for the following $k$ steps. Conditioned on this goal, the executor acts at every environment step:
\begin{equation}
a_t \sim
\pi_{\mathrm{lo}}\!\left(\cdot\mid o_t,g_t\right).
\label{eq:executor}
\end{equation}
\textcolor{black}{In practical terms, the hierarchy first maps a noisy observation to a subnet-level defensive goal and then maps that goal, together with the current observation, to a concrete defensive action.} This fixed-duration hierarchy induces a temporal abstraction in which planner decisions are separated by $k$ primitive transitions, with discount $\gamma^k$ between successive planner decisions. Defense quality is defined by the episodic discounted return
\begin{equation}
J =
\mathbb{E}\!\left[
\sum_{t=0}^{T-1}\gamma^t r_t
\right],
\qquad
r_t=R(s_t,a_t,s_{t+1}),
\label{eq:objective}
\end{equation}
over an episode of length $T$. RL controllers are trained to maximize \eqref{eq:objective}, whereas frozen LLM controllers receive no task-specific parameter updates and are evaluated under the same environment reward.

\subsection{Reward Structure}
\label{sec:reward}
All controller configurations operate under the same defensive objective. The reward is additively decomposed into four terms:

\begin{equation}
r_t =
\underbrace{-\!\!\sum_{c\in\mathcal{C}_t}\sigma(c)}_{\text{attacker progress}}
+
\underbrace{\sum_{b\in\mathcal{B}_t}\beta(b)}_{\text{defensive benefit}}
-
\underbrace{\kappa(a_t)}_{\text{action cost}}
-
\underbrace{\phi(a_t,s_t)}_{\text{collateral disruption}},
\label{eq:reward}
\end{equation}
where $\mathcal{C}_t$ contains the successful attacker sub-actions realized during the $t\!\to\!t{+}1$ transition and $\sigma$ weights them according to kill-chain severity, increasing from reconnaissance to the terminal impact action. $\mathcal{B}_t$ contains productive defensive outcomes, such as diverting the attacker into a decoy, detecting or containing it, or blocking lateral movement, with corresponding benefit $\beta$. The term $\kappa(a_t)$ captures the cost of defensive actions, including immediate and capped recurring costs, while $\phi(a_t,s_t)$ penalizes unnecessary disruption, such as isolating an uncompromised host. For clarity, we omit from \eqref{eq:reward} a zone-matching scale factor used by the implementation when the defender acts outside the attacker's subnet. The values of $\sigma$, $\beta$, $\kappa$, and $\phi$ are provided in Section~\ref{sec:setup-env}. The same reward is used across RL+RL, LLM+RL, and LLM+LLM configurations, so all controllers are evaluated against a common operational objective.

\subsection{Controller Realizations}
\label{sec:controllers}
The planner and executor in \eqref{eq:planner}-\eqref{eq:executor} expose common decision interfaces but do not prescribe how those decisions are produced. We instantiate each role using either a trained RL policy or a frozen, zero-shot LLM. This separation allows controller type to be varied without changing the hierarchical decomposition, environment dynamics, or reward.

\subsection{Experimental Instantiations}
\label{sec:instantiations}
The controller-agnostic formulation allows us to vary how LLM capability is introduced into the hierarchy while keeping the decision process and interfaces fixed.

\begin{itemize}
\item RL+RL is the hierarchical RL baseline, in which both planner and executor are trained with PPO. This configuration represents the conventional learned hierarchy and provides the reference against which retraining-free LLM control is evaluated.
\item LLM+RL replaces only the planner with a frozen, zero-shot LLM while retaining the trained PPO executor. This configuration tests the natural planner-first hypothesis motivated by the hierarchy, as described in the Introduction section.
\item LLM+LLM extends frozen LLM control from planning to execution, replacing the PPO executor with an LLM controller as well. Comparing LLM+RL with LLM+LLM therefore measures what changes when LLM control is extended from strategic targeting to tactical action selection while retaining LLM-based planning.
\end{itemize}

All configurations share the same underlying decision process $\mathcal{M}$, reward~\eqref{eq:reward}, goal set $\mathcal{G}$, planner horizon $k$, and environment interfaces. %Thus, the comparison isolates the effect of changing the controller composition under a common experimental protocol.
Thus, the configurations \textcolor{black}{hold the decision process and environment interfaces fixed while varying controller composition under a common experimental protocol.}

\section{Experimental Setup}
\label{sec:setup}
\subsection{Cyberwheel Environment and Threat Model}
\label{sec:setup-env}
We instantiate the hierarchical defender in Cyberwheel~\cite{cyberwheel}, a high-fidelity cyber defense environment developed by the U.S. Department of Energy's Oak Ridge National Laboratory. Cyberwheel represents an enterprise network as a graph of hosts organized into subnets and connected through routers, routing rules, and firewall policies. We evaluate three network scales: small (15 hosts), medium (100 hosts), and large (1,010 hosts).

\emph{Attacker.} The red agent is scripted and executes an Atomic Red Team-based attack sequence mapped to MITRE ATT\&CK techniques~\cite{atomicredteam, attack}. Under the server downtime objective, the attacker begins from a randomly selected entry host and progresses toward a server through the ordered sequence \emph{pingsweep $\rightarrow$ portscan $\rightarrow$ discovery $\rightarrow$ lateral movement $\rightarrow$ privilege escalation $\rightarrow$ impact}. The terminal impact action disrupts the target service and represents the attacker's primary objective. Consistent with the common reward defined in Section~\ref{sec:reward}, successful attacker actions are weighted according to kill-chain severity:
\begin{equation}
\sigma =
\{1,\,2,\,5,\,10,\,20,\,50\},
\end{equation}
corresponding respectively to pingsweep, portscan, discovery, lateral movement, privilege escalation, and impact.

\emph{Observations.} The defender receives detector-derived observations rather than the underlying network state. For each host, the observation encodes the current alert, decayed alert history, defensive state indicators such as host isolation and decoy deployment, and an inferred attack stage index ranging from 0 (no observed activity) to 6 (impact). These host-level features are augmented with subnet-level aggregates describing current alert volume, historical alert activity, and trend, together with the number of deployed decoys.

Consistent with the planner-executor decomposition in Section~\ref{sec:framework}, the planner receives a subnet-level view $o^{\mathrm{hi}}$ for strategic targeting. The executor acts at every environment step using the current observation together with a one-hot representation of the selected goal $g_t$. Thus, the planner determines \emph{where} defensive attention should be directed, while the executor determines \emph{which action and target} should implement that decision.

\emph{Actions.} The primitive defensive action set is
\begin{equation}
\mathcal{U} =
\{\textsc{Deploy Decoy},\,
  \textsc{Isolate Host},\,
  \textsc{Nothing}\}.
\end{equation}
\textsc{Deploy Decoy} places a decoy within the selected subnet to detect or divert the attacker, \textsc{Isolate Host} removes a host from normal network connectivity to contain lateral movement, and \textsc{Nothing} leaves the current configuration unchanged. Host isolation incurs an immediate action cost of $-1$ and a recurring cost of $-0.1$, whereas decoy deployment incurs no immediate cost but a small recurring cost of $-0.1$, with recurring costs capped at $-1.5$ per step. Successful defensive outcomes receive bonuses of $+15$ for diverting the attacker into a decoy or blocking lateral movement, $+10$ for catching the attacker, and $+25$ for protecting a critical asset, while unnecessary disruption, such as isolating an uncompromised host, incurs a penalty of $-10$. These terms instantiate the common reward defined in Section~\ref{sec:reward}. The planner horizon is fixed at $k=5$, so each selected subnet remains the active goal for five environment steps.

\subsection{Controller Instantiation and LLM Models}
\label{sec:setup-controllers}
The three configurations introduced in Section~\ref{sec:instantiations} differ only in how the planner and executor controllers are instantiated. For each network scale, the RL+RL baseline uses PPO policies trained specifically for that network under the same reward, observation, action, and hierarchical interfaces used by the LLM-based variants. Accordingly, the RL baseline is retrained as network scale changes rather than transferred unchanged across the three environments. \textcolor{black}{The RL and LLM controllers therefore represent different training regimes. Our comparison is intended to evaluate their deployment behavior under common defense interfaces, rather than to provide a matched comparison of training data, compute, or sample efficiency.}

We evaluate six LLMs selected to vary two dimensions, model scale and cybersecurity specialization, relevant to our research questions. The general purpose models are Llama-3.2-3B, Qwen3-8B, Qwen3-32B, and Llama-3.3-70B. We additionally evaluate Foundation-Sec-8B \cite{foundationsec} and Trendyol-Cybersecurity-70B \cite{trendyol} as cybersecurity-specialized models, providing specialized comparators at the 8B and 70B scales. For Qwen3-8B and Qwen3-32B, the models' explicit reasoning/thinking mode is enabled at the planning level.

Each LLM receives a textual rendering of the observation appropriate to its role and returns a structured JSON decision that is parsed into the corresponding planner goal or executor action. All models are served using Q4\_K\_M quantization through Ollama on a single NVIDIA H100 GPU with low-temperature decoding. No gradients or task-specific parameter updates are applied to any LLM.

\subsection{Evaluation Protocol}
\label{sec:protocol}
To preserve comparability across controller configurations, all evaluations use a common global seed so that environment stochasticity and scripted attacker behavior are matched across conditions. Each configuration is evaluated over 20 episodes of 100 environment steps. \textcolor{black}{The common seed supports controlled paired comparisons across configurations, but the present evaluation does not characterize variability across independent random seeds.} The executor uses deterministic (argmax) decoding during evaluation.

Across all LLM experiments, every generated decision is successfully parsed into a valid planner goal or executor action, with no heuristic fallback required. Thus, the reported outcomes reflect the decisions produced directly by the evaluated controllers rather than recovery rules for malformed outputs.

The protocol is designed around two complementary comparisons. First, performance across the small, medium, and large networks tests whether a frozen LLM can retain defensive effectiveness as network scale changes without model retraining. Second, the comparison between LLM+RL and LLM+LLM tests how performance changes when frozen LLM control is extended from strategic planning to tactical execution. RL+RL provides the trained hierarchical reference at each network scale.

\subsection{Evaluation Metrics}
\label{sec:metrics}
We evaluate defensive performance using three step normalized rates. Let $c_t^{\mathrm{lat}}$ indicate a successful red lateral-movement action at step $t$, $c_t^{\mathrm{imp}}$ indicate a successful terminal impact action, and $d_t$ indicate that the red action either fails or is diverted into a decoy. For an episode of length $T$, we define the following metrics:

\begin{itemize}
\item Compromise rate measures the frequency with which the attacker successfully establishes a new foothold through lateral movement. It is interpreted as a step normalized attack progression rate.

\begin{equation}
\mathrm{Compromise}
=
\frac{100}{T}
\sum_{t=0}^{T-1} c_t^{\mathrm{lat}}
\label{eq:compromise}
\end{equation}
\item Impact rate measures how frequently the attacker reaches and successfully executes its terminal kill-chain objective.

\begin{equation}
\mathrm{Impact}
=
\frac{100}{T}
\sum_{t=0}^{T-1} c_t^{\mathrm{imp}}
\label{eq:impact}
\end{equation}
Because impact represents the attacker's most severe outcome and ultimate objective, we treat it as the primary security outcome metric. Lower compromise and impact rates indicate stronger defense.

\item Defense rate is a passive success measure that counts steps in which the red action fails for any reason or is diverted into a decoy. Because this measure can include intrinsic attacker failure rather than an explicit defender block, we use it as a supporting indicator rather than as the primary measure of defender effectiveness; higher values are better.

\begin{equation}
\mathrm{Defense}
=
\frac{100}{T}
\sum_{t=0}^{T-1} d_t
\label{eq:defense}
\end{equation}
In Equation~\ref{eq:defense}, $d_t=1$ when the red action fails or is diverted into a decoy, and $d_t=0$ otherwise.
\end{itemize}

\section{Results}
\label{sec:results}
Table~\ref{tab:results} reports compromise, impact, and defense rates for the various configurations across network scales.

\subsection{\textcolor{black}{RL Baseline Performance Across Network Scale}}%\subsection{RL Performance Degrades with Network Scale}
The trained RL+RL baseline becomes less effective as network size increases despite being trained separately for each scale. Compromise rises from 5.32\% on the small network to 14.88\% and 17.95\% on the medium and large networks, respectively. Impact follows the same pattern, increasing from 1.22\% to 5.10\% and 15.20\%. %These results indicate a scale dependent performance challenge that motivates retraining-free LLM control.
\textcolor{black}{These results establish the scale dependent behavior of the trained RL reference against which the frozen LLM configurations are evaluated.}

\subsection{LLM Planning Alone Provides Limited Gains}
Replacing only the planner with a frozen LLM improves performance in several settings but does not remove the degradation at larger scales. On the large network, the LLM+RL configurations remain clustered between 10.75\% and 13.05\% compromise for all models except the 3B model, while impact remains between 8.40\% and 10.85\%. Increasing planner model size therefore produces relatively limited additional benefit when tactical execution remains with the fixed PPO executor.

\subsection{Extending LLM Control to Execution Enables Strong Cross-Scale Performance}

The largest gains emerge when capable LLMs control both levels of the hierarchy. Llama-3.3-70B achieves compromise rates of 1.15\%, 1.40\%, and 1.00\% across the small, medium and large networks, respectively, while impact remains at 0.10\%, 0.05\%, and 0.00\%. Qwen3-32B shows a similar pattern, with compromise at or below 2.25\% and impact at or below 0.60\% across all three scales. \textcolor{black}{Operationally, the same frozen LLM weights achieve these results across all network sizes without task-specific retraining, while the RL reference is trained separately at each scale.} Comparing LLM+RL with LLM+LLM further shows the effect of extending LLM control to tactical execution. For Llama-3.3-70B on the large network, compromise decreases from 11.75\% to 1.00\% and impact from 9.25\% to 0.00\%. For Qwen3-32B, compromise similarly decreases from 11.75\% to 1.15\%. Thus, planner-only integration does not capture the full benefit observed when a capable LLM also performs tactical action selection.

Notably, LLM control does not improve performance universally. Llama-3.2-3B underperforms the RL baseline at several scales, while Trendyol-Cybersecurity-70B performs well on the small network but degrades on the large network. Conversely, Foundation-Sec-8B outperforms general-purpose Qwen3-8B at the largest scale. These mixed results suggest that cybersecurity specialization can help at matched scale but does not guarantee cross-scale robustness.

\begin{table}[t]
\caption{Defense performance across network scales. C = Compromise (\%), I = Impact (\%), and D = Defense (\%). Bold indicates the best result within each controller configuration. $^\dagger$ denotes a cybersecurity-specialized model.}
\label{tab:results}
\centering
\setlength{\tabcolsep}{2.3pt}
\scriptsize
\begin{tabular}{lccc|ccc}
\toprule
& \multicolumn{3}{c|}{\textbf{LLM+RL}}
& \multicolumn{3}{c}{\textbf{LLM+LLM}} \\
\textbf{Model} & C & I & D & C & I & D \\
\midrule

\multicolumn{7}{l}{\textit{Small network (with RL+RL baseline C=5.32, I=1.22, D=75.52)}} \\
Llama-3.2-3B
& 5.45 & 10.25 & 64.40 & 10.60 & 4.45 & 68.25 \\
Qwen3-8B
& 6.00 & 3.65 & 70.45 & 7.10 & 3.80 & 75.55 \\
Foundation-Sec-8B$^\dagger$
& \textbf{4.35} & \textbf{0.25} & \textbf{78.00} & 6.35 & 0.15 & 72.30 \\
Qwen3-32B
& 4.80 & 3.50 & 73.60 & 2.10 & 0.60 & 90.70 \\
Trendyol-70B$^\dagger$
& 6.90 & 13.65 & 56.10 & 1.30 & \textbf{0.05} & \textbf{92.45} \\
Llama-3.3-70B
& 7.15 & 16.45 & 53.10 & \textbf{1.15} & 0.10 & 90.90 \\

\midrule
\multicolumn{7}{l}{\textit{Medium network (with RL+RL baseline C=14.88, I=5.10, D=68.38)}} \\
Llama-3.2-3B
& 14.75 & 5.15 & 30.05 & 17.85 & 5.75 & 22.10 \\
Qwen3-8B
& 13.35 & 3.85 & 37.30 & 6.95 & 3.90 & 65.20 \\
Foundation-Sec-8B$^\dagger$
& 11.35 & 2.70 & 47.05 & 10.20 & 3.20 & 50.60 \\
Qwen3-32B
& 10.45 & \textbf{2.25} & 48.45 & 2.25 & 0.25 & 84.70 \\
Trendyol-70B$^\dagger$
& \textbf{9.20} & 2.40 & \textbf{51.55} & 8.70 & 2.80 & 57.80 \\
Llama-3.3-70B
& 11.95 & 3.20 & 41.50 & \textbf{1.40} & \textbf{0.05} & \textbf{87.00} \\

\midrule
\multicolumn{7}{l}{\textit{Large network (with RL+RL baseline C=17.95, I=15.20, D=50.01)}} \\
Llama-3.2-3B
& 11.15 & 9.10 & 28.40 & 17.00 & 14.10 & 2.20 \\
Qwen3-8B
& 11.75 & 9.25 & 25.00 & 4.60 & 2.95 & 65.90 \\
Foundation-Sec-8B$^\dagger$
& \textbf{10.75} & \textbf{8.40} & \textbf{32.30} & 1.85 & 0.85 & 82.30 \\
Qwen3-32B
& 11.75 & 9.25 & 25.00 & 1.15 & 0.10 & 83.10 \\
Trendyol-70B$^\dagger$
& 13.05 & 10.85 & 20.05 & 14.45 & 11.60 & 12.95 \\
Llama-3.3-70B
& 11.75 & 9.25 & 25.00 & \textbf{1.00} & \textbf{0.00} & \textbf{87.95} \\

\bottomrule
\end{tabular}
\end{table}

\section{Conclusion and Future Work}
\label{sec:conclusion}
We investigated whether frozen, zero-shot LLMs can 
\textcolor{black}{provide effective hierarchical cyber defense across network scales without task-specific retraining} and how performance changes as LLM control extends from strategic planning to tactical execution.
%reduce the retraining dependence of hierarchical RL for autonomous cyber defense and how performance changes as LLM control extends from strategic planning to tactical execution.
Across Cyberwheel networks, planner-only LLM integration provides limited gains at larger scales, whereas capable LLMs controlling both levels achieve strong cross-scale performance. Smaller and some cybersecurity-specialized models degrade as networks grow, indicating that domain specialization does not guarantee scale generalization. Adding pretrained reasoning only at the top of a hierarchical defender may be insufficient; effective tactical execution is important for translating strategic decisions into defensive performance. %\textcolor{red}{These findings demonstrate cross-scale reuse of pretrained LLM controllers under the evaluated conditions, rather than a matched comparison of the learning efficiency or training requirements of LLMs and RL.}

Several questions remain. First, evaluating the complementary RL+LLM configuration would better isolate the contributions of planning and execution. Second, analyzing planner and executor decision traces could explain why some frozen LLMs generalize across scales while others degrade. %Finally, future work should test generalization beyond network size, including unseen topologies, attacker strategies, and multi-agent enterprise environments.
Finally, future work should \textcolor{black}{extend evaluation across multiple independent seeds and} test generalization to unseen topologies, attacker strategies, and multi-agent enterprise environments.
\vspace{-0.25em}
\section*{ACKNOWLEDGMENTS}
\vspace{-0.25em}
This work was supported in part by the U.S. Military Academy (USMA) under Cooperative Agreement No. W911NF-25-2-0008 and the Defense Advanced Research Projects Agency (DARPA) under Cooperative Agreement No. HR0011-24-2-0004. The views and conclusions expressed in this paper are those of the authors and do not reflect the official policy or position of USMA, U.S. Army, U.S. Department of War, or U.S. Government.

\end{document}